# Orbital-selective Kondo-lattice and enigmatic *f*-electrons emerging from inside the antiferromagnetic phase of a heavy fermion

**Authors**

Ioannis Giannakis[1], Justin Leshen[1], Mariam Kavai[1], Sheng Ran[2,3], Chang-Jong Kang[4], Shanta R. Saha[2,3], Y. Zhao[2,5], Z. Xu[2], J. W. Lynn[2], Lin Miao[6,7], L. Andrew Wray[6], Gabriel Kotliar[4,8], Nicholas P. Butch[2,3], and Pegor Aynajian[1]*

**Affiliations**

[1]Department of Physics, Applied Physics and Astronomy, Binghamton University, Binghamton, New York 13902, USA

[2]NIST Center for Neutron Research, National Institute of Standards and Technology, Gaithersburg, Maryland 20899, USA

[3]Center for Nanophysics and Advanced Materials, Department of Physics, University of Maryland, College Park, Maryland 20742, USA

[4]Department of Physics and Astronomy, Rutgers University, New Jersey 08854, USA

[5]Department of Materials Science and Engineering, University of Maryland, College Park, Maryland 20742, USA

[6]Department of Physics, New York University, New York, New York 10003, USA

[7]Advanced Light Source, Lawrence Berkeley National Laboratory, Berkeley, CA 94720, USA

[8]Condensed Matter Physics and Materials Science Department, Brookhaven National Laboratory, Upton, New York 11973, USA

* To whom correspondence should be addressed: aynajian@binghamton.edu

**Abstract**

**Novel electronic phenomena frequently form in heavy fermions as a consequence of the mutual nature of localization and itineracy of *f*-electrons. On the magnetically ordered side of the heavy fermion phase diagram, *f*-moments are expected to be localized and decoupled**

from the Fermi surface. It remains ambiguous whether a Kondo-lattice can develop inside the magnetically ordered phase. Using spectroscopic imaging with the scanning tunneling microscope, complemented by neutron scattering, x-ray absorption spectroscopy, and dynamical mean-field theory, we probe the electronic states in the antiferromagnetic $USb_2$ as a function of temperature. We visualize a large gap in the antiferromagnetic phase at high temperatures (T < $T_N$ ~ 200 K) within which Kondo-hybridization gradually develops below $T^{coh}$ ~ 80 K. Our dynamical mean-field theory calculations indicate the antiferromagnetism and Kondo-lattice to reside predominantly on different *f*-orbitals, promoting orbital-selectivity as a new conception into how these two phenomena coexist in heavy fermions. Finally, at T* = 45 K we discover a novel 1st order-like electronic transition through the abrupt emergence of non-trivial 5*f* electronic states that may share some resemblance to the "hidden order" phase of $URu_2Si_2$.

## Introduction

The dual nature of the *f*-electronic wavefunction in heavy fermions drives fascinating electronic behaviors from exotic orders(*1*), to quantum criticality(*2*) and emergent superconductivity(*3–5*). The stage is set at relatively high temperatures by the local Kondo-interaction of the *f*-moments. As temperature is lowered below the so-called coherence temperature, hybridization of these *f*-moments with conduction electrons drives the hitherto localized *f*-electrons into the Fermi sea, enlarging the Fermi surface(*6*). Whether the Kondo quenching is complete or not depends on its competition with the Ruderman-Kittel-Kasuya-Yosida (RKKY) interaction that tends to stabilize long range magnetic order. Theoretically, two scenarios, regarding the zero temperature phase diagram of heavy fermions, have been put forward. The first describes the phase diagram with two separate quantum critical points (QCP), one for the onset of long range magnetic order and another for the destruction of the Kondo-lattice. In the second scenario both phenomena vanish simultaneously at a single QCP(*7–11*). Over the past decade, theoretical and experimental efforts have focused on this very matter of understanding which scenario better describes the phase diagram of the different heavy fermion material systems. There may now be consensus that, at least, YbRh$_2$Si$_2$ follows the latter scenario of Kondo destruction at the magnetic QCP(*12*, *13*). Yet a comprehensive picture remains missing. Differentiating these scenarios not only provides an understanding of the complex phase diagram of heavy fermions but describes the nature of the unconventional critical fluctuations that occur on an extended phase-space near the QCP (the quantum critical fan) out of which unconventional superconductivity often emerges.

In 4*f*-electron systems, such as Ce- and Yb-based heavy fermions (typically with a valence close to 4f$^1$ and 4f$^{13}$ leading to 1 *f*-electron and 1 *f*-hole, respectively), the *f*-orbitals are relatively localized at high temperature and the low temperature properties are well described by the Kondo-lattice model. On the other hand, in 5*f*-electron systems, such as U-based heavy fermions (where determining the valence, typically ranging between 5f$^1$ and 5f$^3$, is more challenging), the spatial extent of the 5*f*-orbitals and their hybridization with ligand orbitals (conduction electrons) gives the *f*-electrons an intermediate character between partial itinerancy and partial localization, increasing the complexity of the Kondo-lattice problem. U-based heavy fermion systems therefore provide an ideal platform to probe this rich and complex physics.

USb$_2$ features one of the highest antiferromagnetic (AFM) transition temperatures (T$_N$) in *f*-electron systems with T$_N$ exceeding 200 K(*14*). Angle resolved photoemission (ARPES)(*15*) and de Haas van Alphen (dHvA)(*16*) experiments indicate two-dimensional cylindrical Fermi surface sheets in the AFM phase. Resistivity measurements(*17*) along the c-axis reveal a peak around 80 K, which may resemble a Kondo-coherence-like behavior well below T$_N$, yet its true origin remains unknown. Despite the high AFM transition temperature, specific heat(*16*) indicates a sizable Sommerfeld coefficient of 27 mJ/(mol.K$^2$), further providing indications of low temperature itinerant heavy quasiparticles residing near the Fermi energy (E$_F$). Spectroscopic investigation of USb$_2$, a clean stoichiometric metal with a very large T$_N$, may therefore provide invaluable information on the interplay between magnetic ordering and Kondo-breakdown phenomena. Moreover, the robust antiferromagnetism in USb$_2$, which drives out of strong correlations, has become particularly interesting very recently as it has been shown that it can be tuned towards quantum critical and tricritical points(*18*, *19*), opening new avenues for emerging phenomena.

Over the past several years scanning tunneling microscope (STM) experiments had considerable success in the imaging of heavy fermions and their formation(*20–24*), their transition into hidden order(*20*, *21*, *25*) and into heavy electron superconductivity(*26*, *27*) in U-, Ce-, and Yb-based systems. Yet in all these studied cases, the magnetically ordered phase and the impact of magnetism on heavy fermion and Kondo-lattice formation has not been accessible. When the local *f*-moments order into an AFM phase, can the Kondo-lattice coherence still develop well below $T_N$ or does it simply breakdown? $USb_2$ is particularly selected to answer this question due to the large accessible temperature window below $T_N$ within which the gradual evolution of the electronic states inside the AFM phase can be visualized.

Using spectroscopic imaging with the STM complemented by elastic neutron scattering, we probe signatures of the magnetic order and Kondo hybridization in $USb_2$ as a function of temperature. We find a large gap ($\Delta \sim 60$ meV) in the tunneling density of states of $USb_2$, which we associate to the AFM order, within which Kondo hybridization near $E_F$ develops with decreasing temperature. Its gradual onset near T = 80 K coincides with the broad peak observed in resistivity measurements(*16*, *17*). At low temperatures, we visualize the hybridization of this heavy band with the conduction electrons, altogether providing spectroscopically conclusive evidence of Kondo coherence emerging from deep within the AFM phase. Our dynamical mean field theory (DMFT) calculations and spatial spectroscopic mapping provide a qualitative description of an orbital-selective AFM order and Kondo lattice that reside on different non-degenerate 5*f* orbitals, offering a new conception of how itinerant and localized *f*-electrons can coexist in heavy fermions. Finally, we discover a novel 1st-order-like electronic transition at T* = 45 K through the abrupt emergence of sharp *f*-electronic states below ($E_1$=-20 meV) and above ($E_2$ = +30 meV) $E_F$ with distinct orbital character.

**Results**

Figure 1a,c displays STM topographs of $USb_2$ and Th doped $U_{1-x}Th_xSb_2$ showing square atomic lattice with a spacing of 4.2 Å corresponding to the in-plane lattice constant of $USb_2$.

Occasionally vertical terraces with spacing of multiples of 8.7 Å are observed (Fig.1b) that match the c-axes of the USb$_2$ single crystal. At least six different single crystals (pure, x = 0.3% Th- and x= 0.5% Th-doped) were cleaved in the current study and numerous areas (spanning several tens of microns) on each sample were probed with the STM. Only one kind of surface has been observed in all these trials, indicating that the cleaving occurs at the Sb2 layer (see Fig.1d(*28*)). Any other cleaving plane statistically results in two different exposed surfaces, not observed in the experiment (see SI figure S1).

Fig.2a shows dI/dV spectra measured at various temperatures away from any defects or dopants on a 0.3% Th-doped U$_{1-x}$Th$_x$Sb$_2$. The spectra away from defects are identical (within the experimental resolution) between all the different samples studied, independent of the minute Th-doping (see SI figure S2). At T = 80 K, the spectrum shows a particle-hole symmetric partial gap in the local density of states via the suppression of spectral weight within ± 60 meV and its buildup in the coherence peaks right above. At this temperature, USb$_2$ is deep in the AFM phase and the observed spectral gap is therefore likely the result of the AFM order. To further elaborate, we carry out temperature dependent elastic neutron scattering on the very same samples. Fig.2b displays the temperature dependence of the AFM (1 1 0.5) Bragg peak that onsets at T$_N$ ~ 203 K. The AFM ordering wave vector indicates a doubling of the Brillouin zone along the c-axis (see Fig.1d), folding of the band structure and gapping over 60% of the Fermi surface at 80 K, seen in our STM data (Fig.2a). The multi-orbital nature of the Fermi surface in USb$_2$ makes it difficult to conclusively determine the origin of metallic AFM phase. However, from our neutron scattering data, we calculate a large magnetic moment of 1.9 μ$_B$/U, consistent with previous reports(*14*), indicating a rather local character.

Looking closer at the 80 K spectrum reveals a weak hump-structure within the AFM gap around the energy E$_0$ ~ 5 meV above E$_F$ (Fig.2a). As temperature is lowered to 52 K, the AFM gap deepens, the coherence peaks (above ± 60 meV) sharpen, and the hump at E$_0$ gradually evolves into a pronounced peak-like structure. Quite remarkably, this resonance observed at E$_0$ in USb$_2$ is in analogy with a similar resonance observed in a different U-based heavy fermion, the celebrated $URu_2Si_2$(*21*, *25*). Fig. 2c compares STM spectra from the two material systems ($URu_2Si_2$ spectra taken from ref. 25). This surprising similarity of the hump-structure near the Fermi energy, in rather two different materials, suggests a similar origin of the low-energy physics, which in $URu_2Si_2$ has been attributed to heavy fermion formation above its hidden order phase(*20*, *25*). The energy position of the $E_0$ resonance which is related to the valence of *f*-electrons(*6*), being a few meV above $E_F$ in both U-systems further indicates similar valence in both systems. This is in contrast to $CeCoIn_5$ ($4f^1$), where the resonance is located right at $E_F$(*25*). Yet, how a Kondo lattice develops well below a magnetically ordered phase in $USb_2$ is a puzzle. To provide insights into this rather bizarre observation we use DMFT to compute the temperature-dependent electronic structure of $USb_2$. We find predominantly a $5f^2$ valence of the U electronic states, in good agreement with our x-ray absorption spectroscopy (see SI figure S3). Quite remarkably, these calculations qualitatively predict both the AFM order and the Kondo behavior with orbital selectivity, where the $m_j = \pm 5/2$ and $\pm 3/2$ orbitals are responsible for the RKKY-type AFM order while the $m_j = \pm 1/2$ orbitals lead to the Kondo resonance near $E_F$ (see SI figure S4).

Lowering the temperature further below T* ~ 45 K (a drop of only a few Kelvin) reveal an abrupt change of the spectral structure in our STM experiments, signaling a sudden transition in the electronic density of states of $USb_2$ (Fig.2a). Note that thermal broadening, which can explain the sharpening of the spectra between 80 K and 52 K is far from describing the observed transition at 45 K. The most pronounced feature is the discontinuous formation of two sharp resonances at $E_1$ ~ -20 meV and $E_2$ ~ +30 meV, far below and above $E_F$. Possible artifacts arising from the STM tip-junction as an explanation to the observed behavior can be excluded since the observed results were reproduced on three different samples, using three different STM tips, and in both cases of crossing T* ~ 45 K from below (warming-up) and above (cooling-down) (see SI figure S2). Such a drastic change of the spectral structure with temperature has not been seen in any of the Ce- Yb- or Sm- based heavy fermion systems studied so far by STM or ARPES. While the $E_{1,2}$ resonances emerge abruptly, the resonance at $E_0$, closer to the Fermi energy, continues to evolve rather gradually with decreasing temperature in analogy with signatures of Kondo-lattice formation observed in other heavy fermion systems.

The co-tunneling mechanism from the STM tip to a sharp resonance in the sample's local density of states manifests itself in a Fano-lineshape (*29–31*). To extract their temperature evolution, we fit the spectra in Fig.2a to three distinct Fano-lineshapes centered around $E_{1,2}$ and $E_0$ with a parabolic background (see SI figure S5). The model is an excellent fit to the data shown as thin dashed lines on top of each spectrum (Fig.2a). The extracted linewidths of the resonances at $E_1$ and $E_0$ are displayed in Fig.2d, e (see SI section S5). The heights of the resonances are extracted directly from the data and are displayed in Fig.2f, g (see SI section S5). As seen in the raw data, the resonance height at $E_0$ gradually grows with a logarithmic temperature dependence below the onset temperature of ~ 80K, providing a long-awaited explanation to the broad peak observed in resistivity measurements. Such a behavior is consistent with Kondo lattice formation observed in other heavy fermions(*20–24*). Similarly, the width of the resonance at $E_0$ decreases in the same temperature range and saturates at low temperature to an intrinsic linewidth expressed in half width at half maximum (HWHM) of $\Gamma$ ~ 6 ± 2 meV. We can analyze the temperature broadening of the linewidth using the Kondo impurity model $\Gamma = \sqrt{(\pi k_B T)^2 + 2(\pi k_B T_K)^2}$, which previously has well-captured the temperature dependence in Kondo lattice systems(*20, 24*). A fit (of a single parameter) of the above equation to the data yields a temperature of $T_K$ = 55 ± 20 K, which is related to the intrinsic linewidth of the Kondo resonance at T = 0 K. Due to the valence state, in U-compounds, the Kondo resonance appears a few meV above $E_F$(*21, 25*), whereas in Ce(*26, 27*), Sm(*32–34*), and Yb(*24*) compounds its observed to be right above, at, and right below $E_F$, respectively. The fact that $E_0$ is in the unoccupied side of the Fermi energy, places ARPES experiments, to probe the lineshape of the resonance, at a disadvantage.

In contrast to this $E_0$ resonance, that at $E_1$ (and $E_2$) follows a very unconventional temperature dependence. It's abrupt onsets at T* ~ 45 K closely matches with the sharp anomaly seen in specific heat and the sudden release of entropy (see supplementary information in ref. (*35*)).

Similarly, optical pump probe spectroscopy(*35*) also sees an abrupt change at T* = 45 K in the lifetime of spin excitations, indicating the opening of a new decay channel, whereas low temperature ARPES(*36*) sees a flat band at the same energy and widths as our $E_1$ resonance. The energy separation between the two resonance levels of $\delta_{1-2}$ ~ 50 meV seen in our spectroscopic measurements corresponds to an excitation temperature exceeding 500 K. Neither their onset nor their signature in specific heat can be explained by the thermal population of conventional crystal-electric-field (CEF) levels and Schottky anomaly. All together these observations hint to an electronic transition in the bulk. Its energy being far from $E_F$ ($E_1 \gg k_B T$) also makes it less sensitive to measurements such as resistivity and only the tail of the $E_1$ resonance near $E_F$ will be captured by transport studies. Once again, our DMFT calculations provide an insight into the origin of these *f*-electronic states. As temperature is lowered, DMFT indicates two peaks that form around ± 30 meV of predominantly $m_j = \pm 3/2$ orbital character, qualitatively in agreement with our observation. While neither their sharpness nor their sudden onset are captured by these calculations, their distinct orbital character nevertheless suggest that some form of orbital ordering may be behind the experimentally observed transition.

To gain access to the origin and the energy–momentum structure of these sharp resonances at $E_{1,2}$ and $E_0$, we carry out spectroscopic imaging with the STM(*37*) at T = 8.7 K. To enhance the scattering signal on a rather clean system (See Fig.1a) we introduce Th-substitution in $U_{1-x}Th_xSb_2$. STM topographs, shown in Figs.1c, reveal four sets of defects, three of which correspond to Th atoms replacing the U atoms in three consecutive U-layers (Fig.1e,f,g) and one which corresponds to Th atoms replacing Sb1 (Fig.1h) (see SI figure S1). At these low concentrations (x=0.3% and 0.5%), the thermodynamic properties of $USb_2$ are unchanged. We first look spatially at isolated Th-dopants with atomic resolution (Fig.3a). The local tunneling density of states measured on a (non-magnetic) Th-atom that substitutes a (magnetic) U-atom right below the surface contrasted with one measured on a clean area away from dopants are shown in Fig.3b (high temperature) and Fig.3c (low temperature). At both temperatures, the spectral features associated with the AFM gap and sharp resonances, all exhibit notable suppression near the Th-atoms (Fig.3b, c), indicating that they originate from electronic states with *f*-character. The Th-atoms in this case are well represented by the so-called Kondo-hole picture. In recent years, extensive theoretical work(*38, 39*) has been carried out on modelling the electronic states surrounding a Kondo-hole. The predicted impact of these Kondo-holes is the atomic scale oscillations of the hybridization strength(*38–40*). The experimental linewidth Γ of the Kondo resonance at $E_0$ is a measure of the Kondo hybridization strength (see SI section S6). From sub-atomic-resolution spectroscopic imaging maps near Th-atoms, we extract the spatial variation of the resonance linewidth, energy and asymmetry q-parameter at low temperature (see SI figure S6). The results are displayed in Fig.3d-i. The first notable observation is the spatial extent of the hybridization (particularly that at $E_0$) spanning a surface area of ~ 50-100 U-sites. The size of this local perturbation is similar to what has been observed around Kondo holes in $URu_2Si_2$(*40*) and also comparable to the local hybridization seen around single Kondo impurities in metals (*41*). The second observation is oscillatory patterns forming spatially-anisotropic ripple-like structures around the Th-dopants, theoretically predicted to form near the Kondo-holes(*38, 39*) (see SI figure S6). Note that, theoretically, the oscillation in energy and q-parameter have not been considered. However, since the q-parameter represents the relative tunneling to the resonant states at $E_0$, oscillation in the hybridization strength is therefore expected to drive a similar oscillatory behavior in the q-parameter (see SI section S6). Finally, the different magnitudes of the q-parameter (sensitivity of tunneling to different orbitals(*22*)) for the $E_0$, $E_1$, and $E_2$ resonances as well as their different spatial dependence (refer to SI section S6 and figure S7) suggest that different non-degenerate *f*-orbitals are responsible for the different resonances, supporting the mechanism of orbital-selectivity proposed by our DMFT calculations.

We now move to probe the quasiparticle interference (QPI) off Th-dopants. Elastic scattering and interference of quasiparticles from impurities gives rise to standing waves in the constant energy conductance maps at wavelengths corresponding to $2\pi/q$, where $\mathbf{q} = \mathbf{k}_f - \mathbf{k}_i$ is the momentum transfer between initial ($\mathbf{k}_i$) and final ($\mathbf{k}_f$) states at the same energy. Note that the oscillatory behavior of the resonance linewidth discussed above is different than the QPI at a constant energy. Figure 4a-e display large scale (50nm) conductance maps at selected energies showing QPI near the Th-atoms. The Fourier transform (FT) of these maps shown in Fig.4f-j reflects the momentum structure of the quasiparticles at the corresponding energies. To quantitatively visualize the quasiparticle dispersions, we take linecuts of the energy dependent FTs along the two high symmetry directions (Fig.5). In addition, the different QPI wave vectors observed in the raw FTs are fit to Gaussians at each energy and the results are plotted as orange data-points in Fig.5. The maps reveal three heavy bands at $E_1$, $E_0$ and $E_2$ (the latter only partially observed) and their hybridization with strongly dispersive conduction electron bands. The hybridization is stronger along the [π/a, π/a] direction as compared to [2π/a, 0]. Finally, the effective mass of the electron m*, is extracted from the curvature $\frac{1}{4}\hbar^2 \left[\frac{d^2 E}{dq^2}\right]^{-1}$ of a second order polynomial fit (quadratic band structure) to the QPI bands crossing $E_F$ (red lines in Fig.5a,b). Our findings of m* = (8 ± 2)$m_0$ and (8 ± 6)$m_0$ along (π/a, π/a) and (2π/a, 0) directions, respectively are in good agreement with results from dHvA experiments (*42*).

**Discussion**

Our studies reveal two important findings in the context of heavy fermions. First, the gradual emergence of sharp quasiparticle peak near the Fermi energy ($E_0$ resonance), its Kondo-hole signature around Th-dopants, and its hybridization with conduction electrons provide a comprehensive picture of coherent Kondo-lattice formation developing deep inside the AFM phase ($T_{KL}$ << $T_N$). This conclusion indicates that, in $USb_2$, the coherent Kondo-lattice does not breakdown on the AFM side of the heavy fermion phase diagram, rather it emerges from within. The insensitivity of the AFM order to the formation of Kondo lattice at $T_{coh}$ ~ 80 K in our neutron scattering experiment along with the orbital selectivity provided by DMFT calculations, altogether demonstrate the dual itinerant and localized nature of the *f*-electrons residing on different *f*-orbitals. This picture is different than, for example, the Kondo breakdown scenario observed in $YbRh_2Si_2$, where the Kondo lattice and AFM order are separated by a single QCP.

Second, the discovery of a pair of non-trivial sharp electronic bands, asymmetrically placed above and below the Fermi energy at $E_1$ and $E_2$ emerging abruptly at 45 K indicate a new form of electronic transition. The origin of these flat bands may be suggestive of orbital ordering but the mechanism of their abrupt, most likely 1st order-like transition (Note that no hint of $E_{1,2}$ is observed at 52 K (Fig.2a).), is mysterious and calls for further investigations. Yet, it is tempting to make some speculations about possible connections to the "hidden order" phase of $URu_2Si_2$, which itself is closely connected (1st order phase transition) to an AFM phase(*43*). In $URu_2Si_2$, a pair of quasiparticle bands of the U 5*f*-electrons asymmetrically form very close (few meV) to the Fermi energy(*25*, *44*) and hybridize with conduction electrons in the hidden-order phase(*21*). Though the energy scales involved in the two U-compounds is very different (by an order of magnitude), the overall resemblance with our observation here perhaps raises a possibility of some connection between the two material systems and calls for future investigations.

## Materials and Methods

### Sample growth

Single crystals of $U_{1-x}Th_xSb_2$ were grown out of an Sn flux, using conventional high-temperature solution growth techniques. Small chunks of U, Th and Sn were mixed together according to the ratio U:Th:Sn = 1-x:x:11.5. Single crystals were grown by slowly cooling the U-Th-SN melt from 1100 to 700 °C over 80 h, then decanting off the excess liquid flux. Identification of commercial equipment does not imply recommendation or endorsement by NIST.

### STM

Flat samples of ~(1x1x0.5) mm were glued to a metallic plate using H74F epoxy and a conducting channel to the plate was achieved using H20E silver conducting epoxy. Identification of epoxy products does not imply recommendation or endorsement by NIST. An aluminum post was attached on the exposed surface of the sample using H74F epoxy, oriented along the sample's (001) plane. The samples were cleaved in situ under UHV by knocking the post off, transferred immediately to the microscope head and cooled down to ~ 8.5 K. PtIr tips were used in all experiment and prior to each experiment the tips were prepared on Cu(111) surface that has been treated to several cycles of sputtering with Ar and annealing before placed in the microscope head to cool. The sample and the Cu(111) were placed inside the microscope head next to each other to minimize exposure and preserve the structural integrity of the tip when moving between the two. STM topographies were taken at constant current mode and *dI/dV* measurements were performed using a lock-in amplifier with a reference frequency set at 0.921kHz. The data presented in this paper was collected from 6 successfully cleaved samples equally distributed between the different Th percentage doping (x=0, x=0.3% and x=0.5%). There were negligible differences measured between different cleaves and different areas on a sample. The error bars correspond to uncertainty of one standard deviation.

**Neutron diffraction**

Neutron diffraction measurements were performed (on the same single crystals used in our STM experiments) at the BT-7 thermal triple axis spectrometer at the NIST Center for Neutron Research using a 14.7 meV energy and collimation: open - 25' - sample - 25' - 120'. The magnetic intensity at the (1,0,0.5) peak was compared to the nuclear intensity at the (1,0,1) peak, while the temperature dependence of the (1,1,0.5) peak was used to calculate an order parameter. An $f^2$ magnetic form factor was assumed. The error bars correspond to uncertainty of one standard deviation.

**Dynamical mean field theory**

We have used density functional theory (DFT) plus DMFT(*45*), as implemented in the full-potential linearized augmented plane-wave method(*46*, *47*) to describe the correlation effect on 5f electrons. The correlated U 5f electrons were treated dynamically by the DMFT local self-energy, while all other delocalized spd electrons were treated on the DFT level. The vertex corrected one-crossing approximation(*45*) was adopted as the impurity solver, in which full atomic interaction matrix was taken into account. The Coulomb interaction U = 4.0 eV and the Hund's coupling J = 0.57 eV were used for the DFT+DMFT calculations.

**H2: Supplementary Materials**

Materials and Methods

Fig. S1. Cleaving plane and Th dopants

Fig. S2. T-dependence

Fig. S3. XAS

Fig. S4. DMFT

Fig. S5. dI/dV Fano fit

Fig. S6. Spatial modulation of Kondo hole

Fig. S7. Linecuts of q-maps

Fig. S8. Hybridized bands and DOS

Fig. S9. FT of fit parameters

Fig. S10. FT of conductance maps for x=0.5%

Fig. S11. Raw-symmetrised data comparison

Fig. S12. Linecuts across QPI features

**References and Notes**

1. J. A. Mydosh, P. M. Oppeneer, Hidden order behaviour in $URu_2Si_2$ (A critical review of the status of hidden order in 2014). *Philos. Mag.* **94**, 3642–3662 (2014).

2. O. Stockert, F. Steglich, Unconventional Quantum Criticality in Heavy-Fermion Compounds. *Annu. Rev. Condens. Matter Phys.* **2**, 79–99 (2011).

3. F. Steglich, S. Wirth, Foundations of heavy-fermion superconductivity: lattice Kondo effect and Mott physics. *Reports Prog. Phys.* **79**, 084502 (2016).

4. S. S. Saxena, P. Agarwal, K. Ahilan, F. Grosche, R. Haselwimmer, M. Steiner, E. Pugh, I. Walker, S. Julian, P. Monthoux, G. Lonzarich, A. Huxley, I. Sheikin, D. Braithwaite, J. Flouquet, Superconductivity on the border of itinerant-electron ferromagnetism in $UGe_2$. *Nature*. **406**, 587–592 (2000).

5. D. Aoki, A. Huxley, E. Ressouche, D. Braithwaite, J. Flouquet, J.-P. Brison, E.

Lhotel, C. Paulsen, Coexistence of superconductivity and ferromagnetism in URhGe. *Nature*. **413**, 613–616 (2001).

6. P. Coleman, in *Handbook of Magnetism and Advanced Magnetic Materials*, K. H., P. S., F. M., M. S., I. Zutic, Eds. (John Wiley & Sons, Ltd, Chichester, UK, 2007); http://doi.wiley.com/10.1002/9780470022184.hmm105).

7. Q. Si, S. Rabello, K. Ingersent, J. L. Smith, Locally critical quantum phase transitions in strongly correlated metals. *Nature*. **413**, 804–808 (2001).

8. P. Coleman, C. Pépin, Q. Si, R. Ramazashvili, How do Fermi liquids get heavy and die? *J. Phys. Cond. Matter*. **13**, R723–R738 (2001).

9. J. A. Hertz, Quantum critical phenomena. *Phys. Rev. B*. **14**, 1165–1184 (1976).

10. A. J. Millis, Effect of a nonzero temperature on quantum critical points in itinerant fermion systems. *Phys. Rev. B*. **48**, 7183–7196 (1993).

11. W. Knafo, S. Raymond, P. Lejay, J. Flouquet, Antiferromagnetic criticality at a heavy-fermion quantum phase transition. *Nat. Phys.* **5**, 753–757 (2009).

12. S. Friedemann, T. Westerkamp, M. Brando, N. Oeschler, S. Wirth, P. Gegenwart, C. Krellner, C. Geibel, F. Steglich, Detaching the antiferromagnetic quantum critical point from the Fermi-surface reconstruction in YbRh$_2$Si$_2$. *Nat. Phys.* **5**, 465–469 (2009).

13. S. Paschen, S. Friedemann, S. Wirth, F. Steglich, S. Kirchner, Q. Si, Kondo destruction in heavy fermion quantum criticality and the photoemission spectrum of YbRh2Si2. *J. Magn. Magn. Mater.* **400**, 17–22 (2016).

14. J. Leciejewicz, R. Troc, A. Murasik, A. Zygmunt, Neutron Diffraction Study of Antiferromagnetism in USb$_2$ and UBi$_2$. *Phys. Status Solidi*. **22**, 517–526 (1967).

15. T. Durakiewicz, P. Riseborough, J. Q. Meng, Resolving the multi-gap electronic structure of USb$_2$ with interband self-energy. *J. Electron Spectros. Relat. Phenomena*. **194**, 23–26 (2014).

16. D. Aoki, P. Wi, K. Miyake, N. Watanabe, Y. Inada, R. Settai, Y. Haga, Y. Onuki, P. Wi, K. Miyake, N. Watanabe, Y. Inada, R. Settai, Cylindrical Fermi surfaces formed by a fiat magnetic Brillouin zone in uranium dipnictides. *Philos. Mag. B*. **80**, 1517–1544 (2000).

17. R. Wawryk, Magnetic and transport properties of UBi$_2$ and USb$_2$ single crystals.

*Philos. Mag.* **86**, 1775–1787 (2006).

18.     R. L. Stillwell, I.-L. Liu, N. Harrison, M. Jaime, J. R. Jeffries, N. P. Butch, Tricritical point of the f-electron antiferromagnet $USb_2$ driven by high magnetic fields. *Phys. Rev. B*. **95**, 014414 (2017).

19.     J. R. Jeffries, R. L. Stillwell, S. T. Weir, Y. K. Vohra, N. P. Butch, Emergent ferromagnetism and T-linear scattering in $USb_2$ at high pressure. *Phys. Rev. B*. **93**, 184406 (2016).

20.     P. Aynajian, H. Eduardo, S. Neto, C. V Parker, Y. Huang, A. Pasupathy, J. Mydosh, A. Yazdani, Visualizing the formation of the Kondo lattice and the hidden order in $URu_2Si_2$. *Proc. Natl. Acad. Sci.* **107**, 10383–10388 (2010).

21.     A. R. Schmidt, M. H. Hamidian, P. Wahl, F. Meier, A. V. Balatsky, J. D. Garrett, T. J. Williams, G. M. Luke, J. C. Davis, Imaging the Fano lattice to 'hidden order' transition in $URu_2Si_2$. *Nature*. **465**, 570–576 (2010).

22.     P. Aynajian, E. H. da Silva Neto, A. Gyenis, R. E. Baumbach, J. D. Thompson, Z. Fisk, E. D. Bauer, A. Yazdani, Visualizing heavy fermions emerging in a quantum critical Kondo lattice. *Nature*. **486**, 201–206 (2012).

23.     A. Maldonado, H. Suderow, S. Vieira, D. Aoki, J. Flouquet, Temperature dependent tunneling spectroscopy in the heavy fermion $CeRu_2Si_2$ and in the antiferromagnet $CeRh_2Si_2$. *J. Phys. Condens. Matter*. **24**, 475602 (2012).

24.     S. Ernst, S. Kirchner, C. Krellner, C. Geibel, G. Zwicknagl, F. Steglich, S. Wirth, Emerging local Kondo screening and spatial coherence in the heavy-fermion metal $YbRh_2Si_2$. *Nature*. **474**, 362–366 (2011).

25.     P. Aynajian, H. Eduardo, S. Neto, B. B. Zhou, S. Misra, R. E. Baumbach, Z. Fisk, J. Mydosh, J. D. Thompson, E. D. Bauer, A. Yazdani, Visualizing Heavy Fermion Formation and their Unconventional Superconductivity in f-Electron Materials. *J. Phys. Soc. Japan*. **83**, 061008 (2014).

26.     B. B. Zhou, S. Misra1, E. H. da S. Neto, P. Aynajian, R. E. Baumbach, J. D. Thompson, E. D. Bauer, A. Yazdani, Visualizing nodal heavy fermion superconductivity in $CeCoIn_5$. *Nat. Phys.* **9**, 474–479 (2013).

27. P. Allan, F. Massee, I. Light, B. Qpi, M. P. Allan, F. Massee, D. K. Morr, J. Van Dyke, A. W. Rost, A. P. Mackenzie, C. Petrovic, J. C. Davis, Imaging Cooper pairing of heavy fermions in CeCoIn$_5$. *Nat. Phys.* **9**, 468–473 (2013).

28. K. Momma, F. Izumi, VESTA 3 for three-dimensional visualization of crystal, volumetric and morphology data. *J. Appl. Crystallogr.* **44**, 1272–1276 (2011).

29. M. Maltseva, M. Dzero, P. Coleman, Electron cotunneling into a kondo lattice. *Phys. Rev. Lett.* **103**, 206402 (2009).

30. J. Figgins, D. K. Morr, Differential conductance and quantum interference in kondo systems. *Phys. Rev. Lett.* **104**, 187202 (2010).

31. P. Wölfle, Y. Dubi, A. V. Balatsky, Tunneling into Clean Heavy Fermion Compounds: Origin of the Fano Line Shape. *Phys. Rev. Lett.* **105**, 246401 (2010).

32. Z. Sun, A. Maldonado, W. S. Paz, D. S. Inosov, A. P. Schnyder, J. J. Palacios, N. Y. Shitsevalova, V. B. Filipov, P. Wahl, Observation of a well-defined hybridization gap and in-gap states on the SmB$_6$ (001) surface. *Phys. Rev. B*. **97**, 235107 (2018).

33. W. Ruan, C. Ye, M. Guo, F. Chen, X. Chen, G. Zhang, Y. Wang, Emergence of a Coherent In-Gap State in the SmB$_6$ Kondo Insulator Revealed by Scanning Tunneling Spectroscopy. *Phys. Rev. Lett.* **112**, 136401 (2014).

34. S. Rößler, T.-H. Jang, D.-J. Kim, L. H. Tjeng, Z. Fisk, F. Steglich, S. Wirth, Hybridization gap and Fano resonance in SmB$_6$. *Proc. Natl. Acad. Sci.* **111**, 4798–802 (2014).

35. J. Qi, T. Durakiewicz, S. A. Trugman, J. Zhu, P. S. Riseborough, R. Baumbach, K. Gofryk, J. Meng, J. J. Joyce, A. J. Taylor, R. P. Prasankumar, E. D. Bauer, K. Gofryk, J. Meng, J. J. Joyce, A. J. Taylor, R. P. Prasankumar, Measurement of Two Low-Temperature Energy Gaps in the Electronic Structure of Antiferromagnetic USb$_2$ Using Ultrafast Optical Spectroscopy. *Phys. Rev. Lett.* **111**, 057402 (2013).

36. T. Durakiewicz, P. S. Riseborough, C. G. Olson, J. J. Joyce, P. M. Oppeneer, S. Elgazzar, E. D. Bauer, J. L. Sarrao, E. Guziewicz, D. P. Moore, M. T. Butterfield, K. S. Graham, Observation of a kink in the dispersion of f-electrons. *EPL (Europhysics Lett.* **84**, 37003 (2008).

37. A. Yazdani, H. Eduardo, S. Neto, P. Aynajian, Spectroscopic Imaging of Strongly Correlated Electronic States. *Annu. Rev. Condens. Matter Phys.* **7**, 11–33 (2016).

38. J. Figgins, D. K. Morr, Defects in heavy-fermion materials: Unveiling strong correlations in real space. *Phys. Rev. Lett.* **107**, 066401 (2011).

39. J. X. Zhu, J. P. Julien, Y. Dubi, A. V. Balatsky, Local electronic structure and fano interference in tunneling into a kondo hole system. *Phys. Rev. Lett.* **108**, 186401 (2012).

40. M. H. Hamidian, A. R. Schmidt, I. A. Firmo, M. P. Allan, P. Bradley, J. D. Garrett, T. J. Williams, G. M. Luke, Y. Dubi, A. V Balatsky, J. C. Davis, How Kondo-holes create intense nanoscale heavy-fermion hybridization disorder. *Proc. Natl. Acad. Sci.* **108**, 18233–18237 (2011).

41. V. Madhavan, W. Chen, T. Jamneala, M. F. Crommie, N. S. Wingreen, Tunneling into a single magnetic atom: spectroscopic evidence of the Kondo resonance. *Science.* **280**, 567–569 (1998).

42. D. Aoki, P. Wiśniewski, K. Miyake, N. Watanabe, Y. Inada, R. Settai, E. Yamamoto, Y. Haga, Y. Onuki, Crystal Growth and Cylindrical Fermi Surfaces of $USb_2$. *J. Phys. Soc. Japan*. **68**, 2182–2185 (1999).

43. E. Hassinger, D. Aoki, F. Bourdarot, G. Knebel, V. Taufour, S. Raymond, A. Villaume, Flouquet, Suppression of hidden order in $URu_2Si_2$ under pressure and restoration in magnetic field. *J. Phys. Conf. Ser.* **251**, 012001 (2010).

44. H.-H. Kung, R. E. Baumbach, E. D. Bauer, V. K. Thorsmølle, W.-L. Zhang, K. Haule, J. A. Mydosh, G. Blumberg, Chirality density wave of the "hidden order" phase in URu2Si2. *Science.* **347**, 1339–1342 (2015).

45. G. Kotliar, S. Y. Savrasov, K. Haule, V. S. Oudovenko, O. Parcollet, C. A. Marianetti, Electronic structure calculations with dynamical mean-field theory. *Rev. Mod. Phys.* **78**, 865–951 (2006).

46. K. Haule, C. H. Yee, K. Kim, Dynamical mean-field theory within the full-potential methods: Electronic structure of CeIrIn5, CeCoIn5, and CeRhIn5. *Phys. Rev. B*. **81**, 195107 (2010).

47. P. Blaha, Wien2k, an Augmented Plane Wave Plus Local Orbitals Program for

Calculating Crystal Properties, edited by K. Schwarz (Technische Universität Wien, Austria, 2001). (2001).

47. L. Miao, R. Basak, S. Ran, Y. Xu, E. Kotta, H. He, J. D. Denlinger, Y. Chuang, Y. Zhao, Z. Xu, J.W. Lynn, J.R. Jeffries, S.R. Saha, I. Giannakis, P. Aynajian, C. Kang, Y. Wang, G. Kotliar, N. P. Butch, L. A. Wray, High temperature singlet-based magnetism from Hund's rule correlations. *Nat. Comm.* **10**, 644(1-8) (2019).

 Acknowledgments

**General**: We thank Piers Coleman, Wei-Cheng Lee, Chris Singh, and Michael Lawler for helpful discussions.

**Funding:** Work at Binghamton University is supported by the U.S. National Science Foundation (NSF) CAREER under award No. DMR-1654482. G.K. and C.-J.K. are supported by DOE BES under Grant No. DE-FG02-99ER45761.

**Author contributions:** I.G., J.L. and M.K. performed the STM measurements. I.G. performed STM data analysis. S.R., Y.Z., J.W.L., Z.X. and I.G. performed the neutron scattering measurements. C-J.K. and G.K. performed DFT+DMFT calculations. L.M. and L.A.W. carried out x-ray absorption spectroscopy. S.R., S.S., and N.B. synthesized and characterized the materials. P.A. wrote the manuscript. All authors commented on the manuscript.

**Competing interests:** The authors declare no competing interests.

**Data and materials availability:** Data is available upon request

**Figures and Tables**

Figure 1

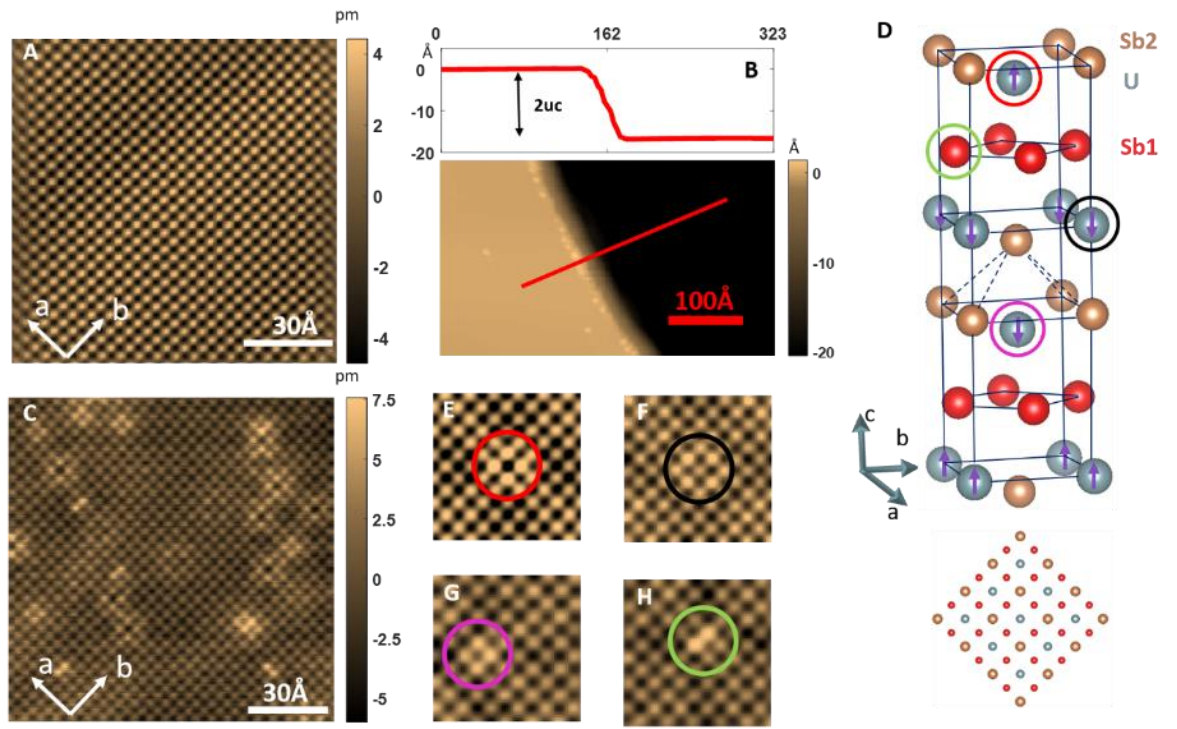

Figure 2

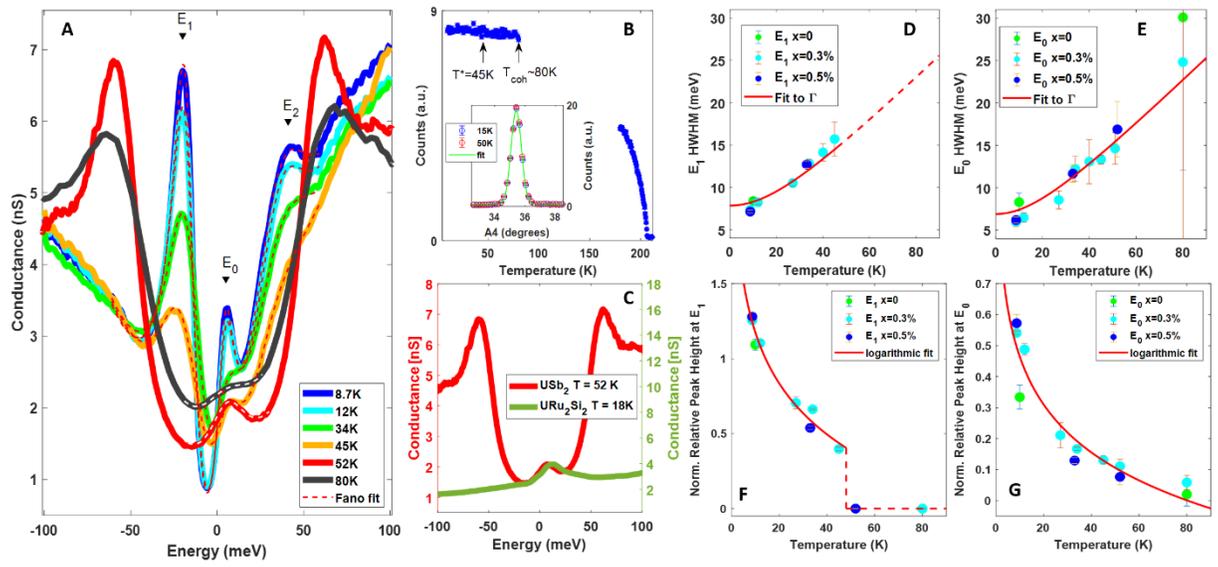

Figure 3

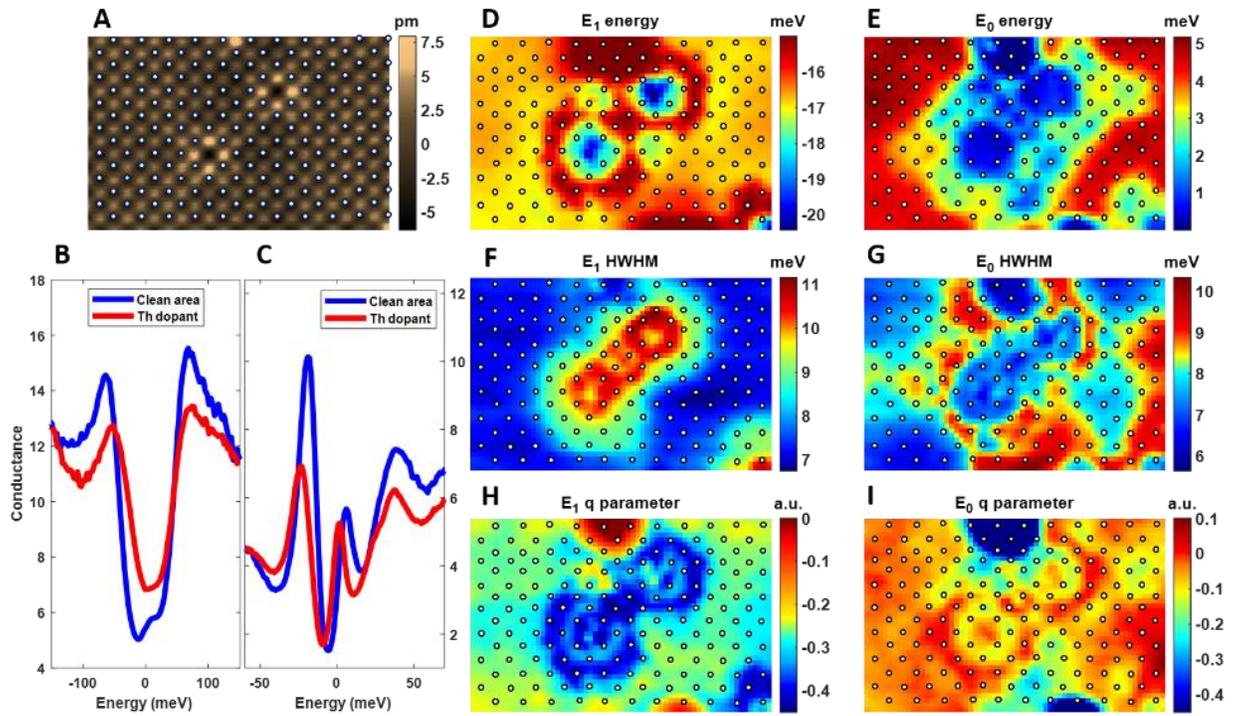

Figure 4

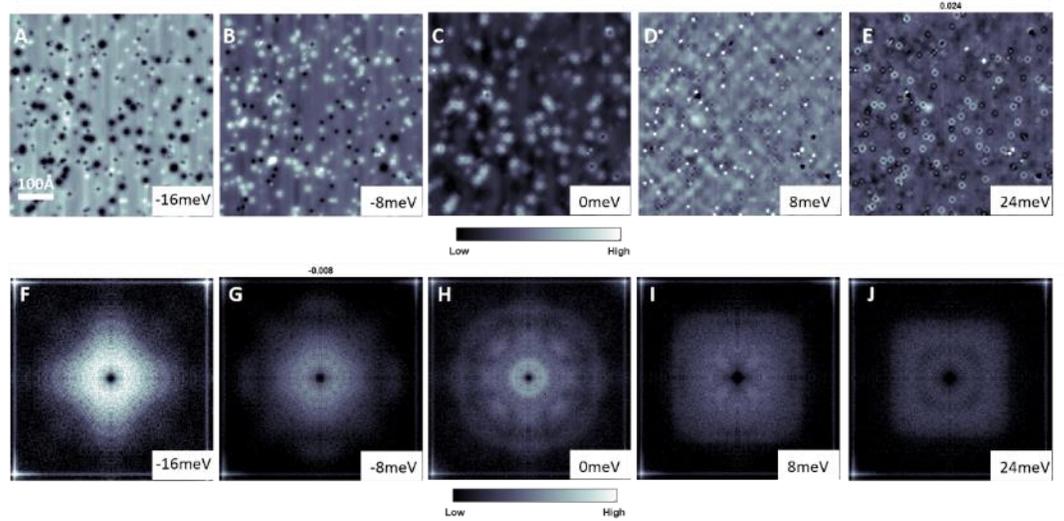

Figure 5

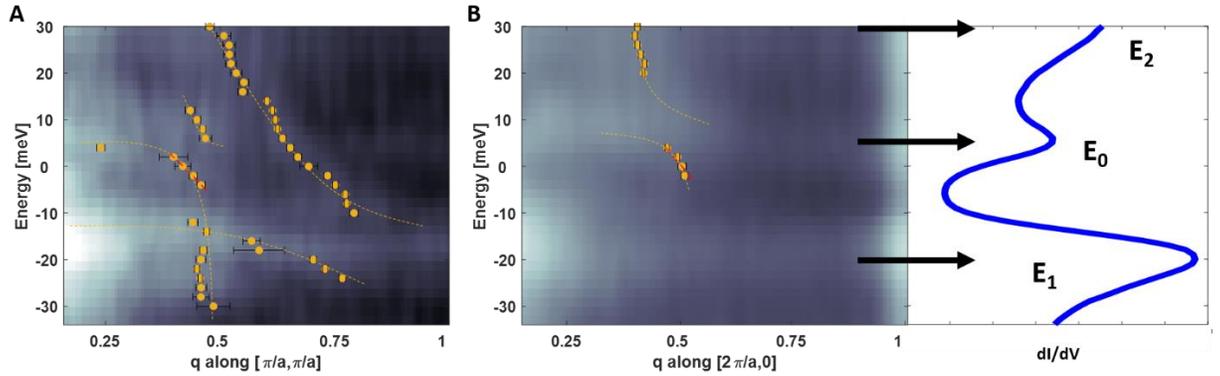

**Fig. 1. Atomic structure of $U_{1-x}Th_xSb_2$.** **(A)** Constant current topographic image on $USb_2$, taken at a set-point bias of 500mV and current of 200pA, showing the atomic surface. **(B)** The topograph shows two terraces separated by 2x c-axis exposing two surfaces with the same chemical termination. The panel above the topograph is a line cut, perpendicular to the step, crossing both terraces. **(C)** Constant current topographic image on the $U_{1-x}Th_xSb_2$ for x=0.5%, taken at a set-point bias of -400mV and current of -1.5nA, showing the atomic surface with four distinct types of defects resulting from Th substitution of U atoms. **(D)** Crystal structure of $USb_2$. The top layer corresponds to the Sb2 layer exposed upon cleaving. The circles mark the position of Th substitutions across two unit cells. The lower panel is a top-down view of the top 3 layers (Sb2-U-Sb1) **(E)-(H)** Th substitutions of U atoms in the $1^{st}$ (red circle), $2^{nd}$ (black circle), $3^{rd}$ (purple circle) and Sb1 (green circle) layer, respectively. The colored circles show the location of the same defects in the compound's crystal structure (Fig. 1d).

**Fig. 2. Electronic states of $U_{1-x}Th_xSb_2$ and their temperature evolution.** **(A)** Evolution of the *dI/dV* spectra ($V_{bias}$ = -400mV, $I_{set\ point}$ = -2nA) from 80K to 8.7K for x=0.3%. At T=80K, a partial gap associated with the AFM state of $USb_2$, as well as a broad hump inside the AFM gap at $E_0$~ 5 meV, can be seen. Below $T^*$~ 45 K, the spectral form changes abruptly revealing two sharp peaks at $E_1$~ -20meV and $E_2$~ 30meV. The dashed lines are fit of the data to three Fano line shapes and a parabolic background. For the spectra above 50K only a single Fano is fitted. **(B)**

Neutron scattering data of the temperature dependence of the AFM Bragg peak (1 1 0.5) that onsets at $T_N \sim 203K$. Inset: θ-2θ of the magnetic Bragg peak at 10K (blue dotted line) and a Gaussian fit (red line). **(C)** STM spectra of $USb_2$ and $URu_2Si_2$ that show similar structure near the Fermi energy. The $URu_2Si_2$ spectrum is taken from ref. 25. It corresponds to the Si-terminated surface, which is analogous to the Sb2-terminated surface of $USb_2$. **(D)-(E)** Temperature dependence of the half width at half maximum (HWHM) of the $E_1$ and $E_0$ resonances, respectively. The lines are fit to $\Gamma = \sqrt{(\pi k_B T)^2 + 2(k_B T_K)^2}$. **(F)-(G)** Temperature dependence of the height of the resonances. The lines are logarithmic fit to the data. The red dashed lines are guides to the eye.

**Fig. 3. Spectroscopic imaging of Kondo-holes. (A)** topographic image showing two Th-dopants and a dimer defect on top-center of the image. **(B)-(C)** dI/dV spectra taken on a clean area (blue) and on a Th-dopant (red) at 80K and 12K, respectively. The spectral features related to the AFM gap and the sharp resonances are suppressed near the Th-dopant. **(D)-(I)** spatial modulation in the vicinity of the Th-dopants of the hybridization energies (d)-(e), the linewidth (f)-(g) and q-parameter (h)-(i) for the peaks at $E_1$ and $E_0$ extracted from fits to the Fano-lineshape.

**Fig. 4. Spectroscopic imaging of the quasiparticle interference. (A)-(E)** Real space conductance maps (-400mV, -2nA) at selected energies for x=0.3% Th doping, showing strong QPI near the Th-dopants. **(F)-(J)** Fourier Transforms (FT) of the same maps revealing rapidly dispersive features of the QPI in the momentum space.

**Fig. 5. Visualizing energy-momentum structure of quasiparticles. (A, B)** Energy dispersion of the quasiparticles scattering along the *(π/a, π/a)* **(A)** and (2π/a,0) **(B)** directions. Three bands corresponding to $E_0$, $E_1$ and $E_2$ appear to hybridize with conduction electrons. The data points are extracted from Gaussian fits to line cuts in the FT (see SI). For (2π/a,0) direction, the dispersion at $E_0$ is very rapid making extraction of the precise wave vector meaningless. The dashed lines are a guide to the eye. The red lines are a quadratic fit to the data points that cross $E_F$. The effective mass is extracted from the curvature $\frac{1}{4}\hbar^2 \left[\frac{d^2E}{dq^2}\right]^{-1}$ of the second order polynomial fit.

# Supplementary Material

## S1. Identifying the cleaving plane and Th-dopants

### Cleaving plane

The accurate identification of the cleaving plane is an important factor in our analysis. We conducted several experiments on the parent compound $USb_2$ as well as the $U_{1-x}Th_xSb_2$ for x=0.3% and 0.5%. Figure S1(A) shows the schematic of the unit cell with red lines representing the possible cleaving planes. The (Sb2-U) plane is a highly unlikely to cleave because of the U-Sb proximity, making it energetically unfavorable. If the sample were to cleave on plane (Sb1-U), then two different surfaces would be exposed with equal probability of encounter, with the Sb1 surface having a lattice constant of factor of $\sqrt{2}$ smaller than the U layer. Contrary, cleaving the sample on plane (Sb2-Sb2) exposes the same chemical surface. On all the samples and the different areas investigated, we encountered only one kind of surface with a lattice spacing corresponding to the lattice constant of the sample. Furthermore, the heights of the steps we observed are consistently an integer multiple of the unit cell's c-axis. These confirm that the sample cleaves on Sb2 layer (see figure 1 of main text) revealing a surface of Sb atoms.

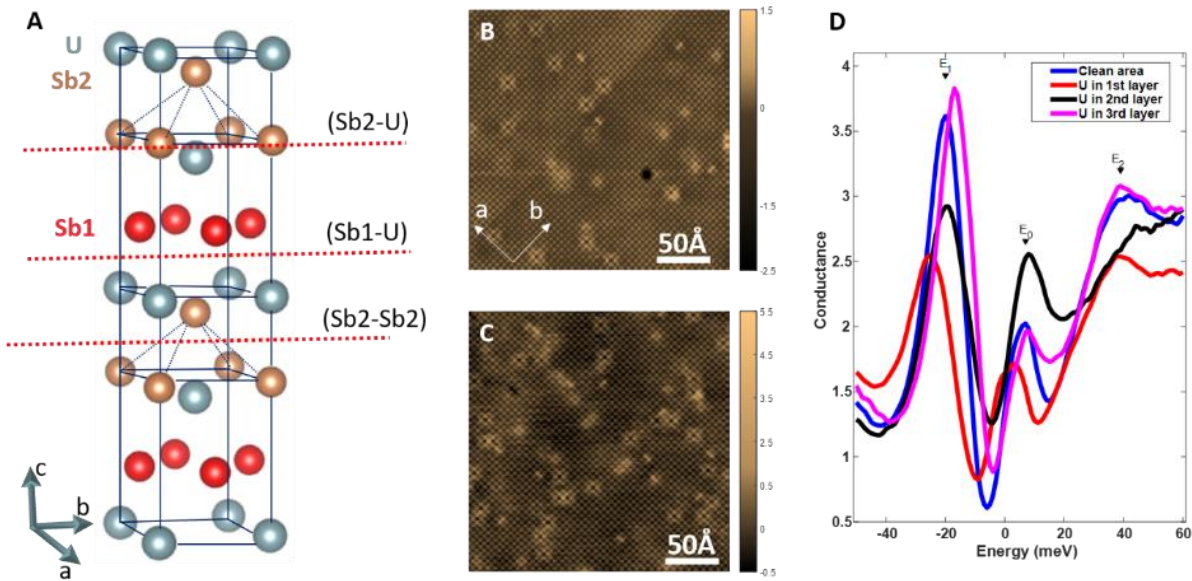

**Fig. S1 Cleaving plane and Th dopants: (A)** Unit cell of $USb_2$. The red lines indicate

the possible cleaving planes. **(B)** x=0.3% and **(C)** x=0.5%. The form of the defects remains unchanged between the two different Th doping percentages. Furthermore, a clear increase of the number of defects is seen as x increases. **(D)** dI/dV spectra taken on the $1^{st}$, $2^{nd}$, $3^{rd}$ layer of Th substituted U sites and on a clean area.

**Th-doping**

The Th-dopants are expected to substitute U sites. From our topographic measurements we observe Th-dopants from three consecutive U-layers in the sample with their signature gradually weakening away from the surface. The defects look either 4-fold symmetric centered between a surface atoms or 4-fold symmetric centered on a surface atom. In our topographies (Fig. S1(B)-(C)) we encounter two kinds of defects that corresponds to the former and one kind that corresponds to the latter description (see figure 1(e-h) of main text) expected from the three U-layers.

Statistical analysis on the number of defects from many areas across different samples shows that the percentage of each of the three 4-fold symmetric defects increases by roughly a factor of two when the sample doping changes from x=0.3% to x=0.5%. For x=0.3%, the concentrations of U in the $1^{st}$, $2^{nd}$ and $3^{rd}$ layer are 0.27±0.07%, 0.33±0.04% and 0.2±0.03% respectively. For x=0.5% the same concentrations increase to 0.61±0.05%, 0.65±0.05% and 0.52±0.04%.
This suggests that all three defects are Th substituting a U site at different locations beneath the surface.
Our reported dopant concentrations, x = 0.3% and x = 0.5%, come from WDS analysis performed on the samples. Transport measurements done on all samples show that within this low doping range, changes in the AFM transition are of the order of 1K. In this context, the introduced dopants serve only as scattering centres.

Further evidence that the Th dopants replace U in three consecutive layers comes from their spectroscopic signature. The different location of the defects along the c-axis results in different spectral form. The peaks $E_1$, $E_0$ and $E_2$ are still present but their intensity gradually changes with respect to the spectra taken on a clean area. The one closest to the surface has the strongest spectroscopic impact and vice versa. Figure S1(D) shows that the suppression of the peaks is greater on the first layer U site. The effect weakens as the Th substitutes a U in the lower layer and eventually disappears completely.

In the topographies we also observe a dimer-shaped defect that follows the orientation of the lattice alternating randomly along a- or b-axes. This form can occur when an atom in the Sb1 plane is substituted. The number of dimers doesn't appear to change between x=0.3% and x=0.5% samples, suggesting they are likely intrinsic defects.

## S2. dI/dV on different Th concentrations and temperature dependence measurements on x=0 and x=0.5%

Spectroscopic dI/dV measurements were carried out on three different samples with different doping levels. Fig. S2 shows the temperature dependent measurements. In all three samples, we observe a high temperature (T > 45 K) partial gap of about 60meV and emergent sharp resonances at low temperatures (T < 45 K). The size of the AFM gap and the position of the resonances are insensitive to the small doping levels as long as the spectra on the doped samples are measured away from the dopants themselves. The dashed lines in the figure show that a kink is observed in the low temperature spectra that coincides with the location of the AFM gap's coherence peaks seen at high T. This suggests that for T<45K that size of the gap remains unchanged. This is further corroborated by the strong kinks observed in the second derivative of the highest and lowest temperature spectra at ±60meV (Fig. S2 D-F), which are located at the same energies as the AFM gap's coherence peaks.

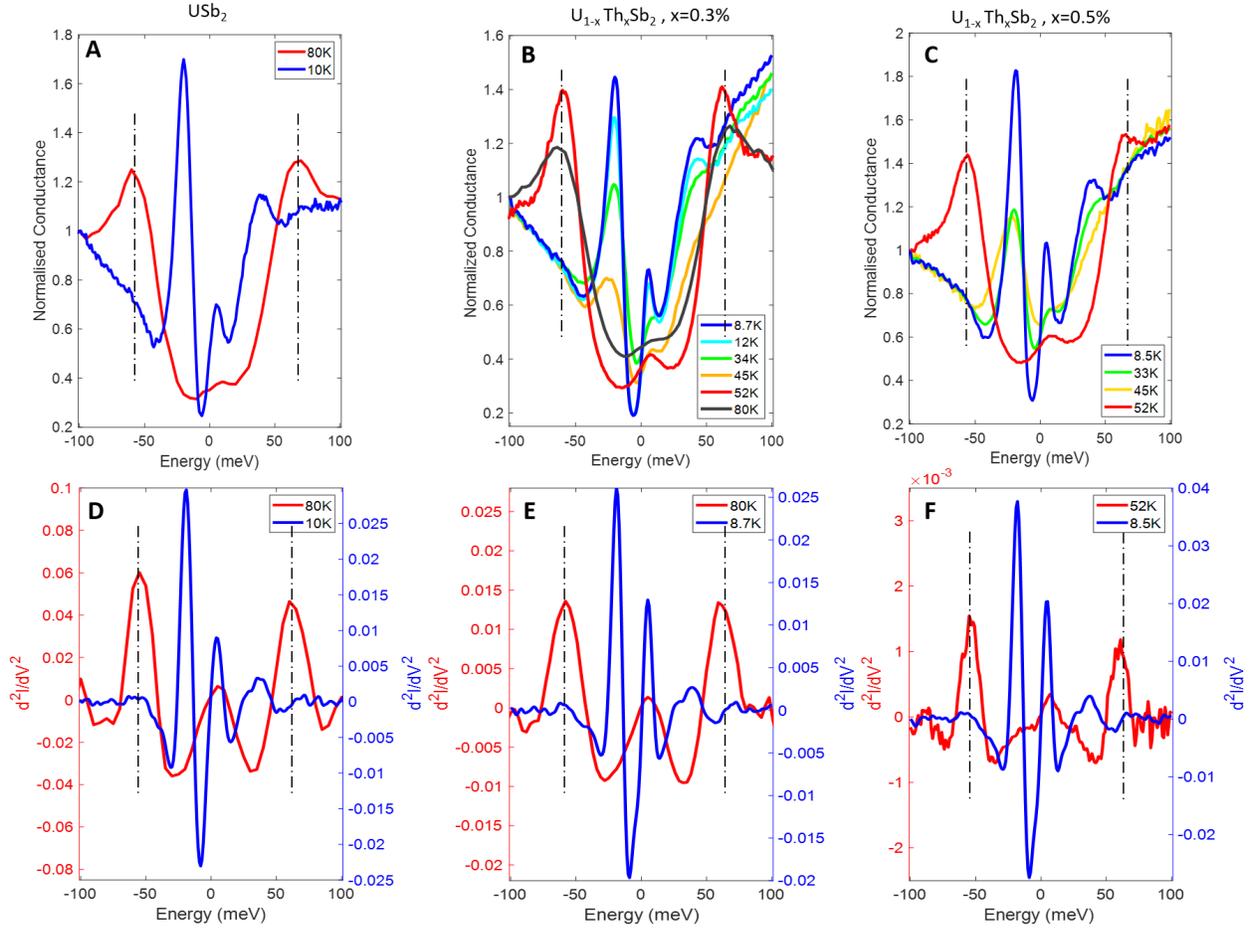

**Fig. S2. T-dependence:** T-dependence measurements on (A) x=0, (B) x=0.3% and (C) x=0.5%. For x=0.5% T* is crossed from below (warm up) while for x=0.3% it is crossed from above (cool down). (D-F) second derivative of the highest and lowest temperature spectra. The second derivative of the low temperature data shows strong kinks at the location of the AFM gap coherence peaks at high temperatures. The size of the AFM gap remains unaffected down to the lowest measured temperature but the spectral weight is transferred inside the gap.

## S3. X-ray absorption spectroscopy (XAS)

The XAS data of $USb_2$ were measured by the total electron yield (TEY) method while for the simulated $5f$ spectra we followed the method in ref. [(48) of the main text] which describes the $5d^{10} 5f^n \rightarrow 5d^9 5f^{n+1}$ X-ray absorption in the dipole approximation. The absorption features of the $USb_2$ spectrum closely match the multiplet simulation for $5f^2$ and a second derivative analysis (SDI) clearly demonstrates the matching spectroscopic features, figure S3.

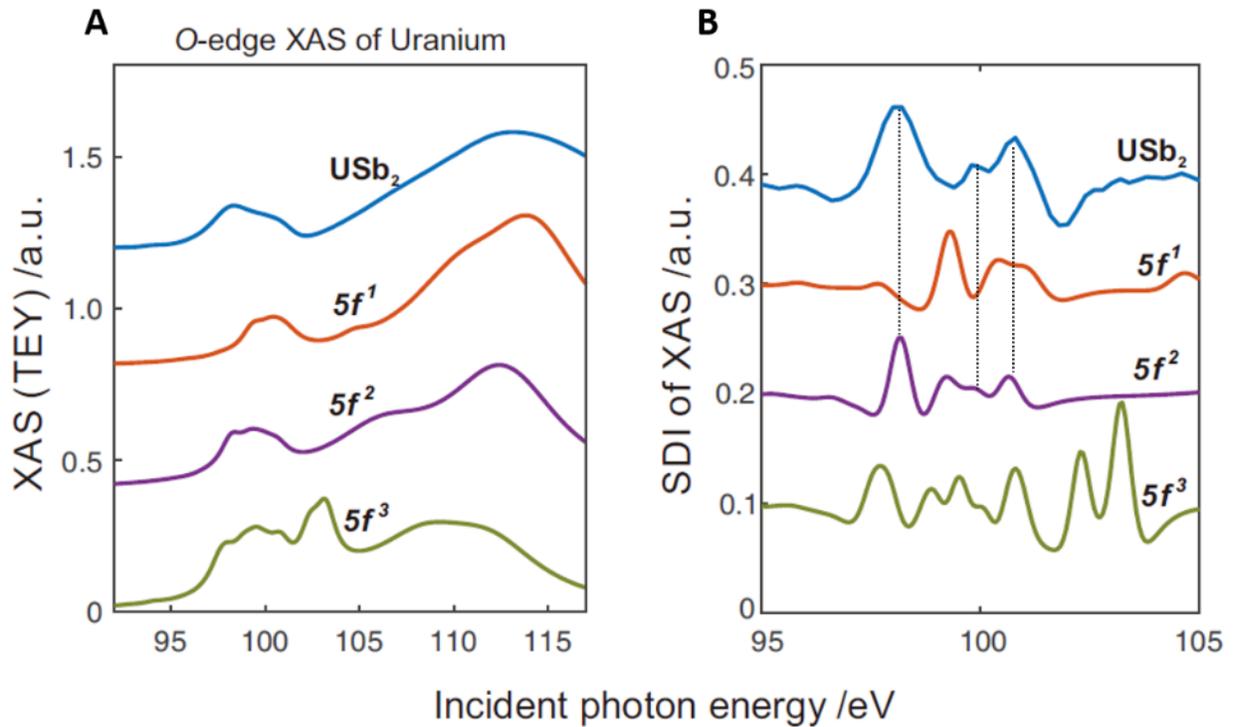

**Fig. S3. XAS: (A)** USB$_2$ XAS spectrum on the uranium O-edge along with the $5f^1$, $5f^2$ and $5f^3$ multiplet simulations. **(B)** A second derivative analysis (SDI) on the curves in (a). The perpendicular lines show the matching spectroscopic figures between the USb$_2$ data and the $5f^2$ simulation.

## S4. Density functional theory and dynamical mean field theory of electronic states of USb$_2$

Density functional theory (DFT) plus dynamical mean field theory (DMFT) was used to study temperature-dependent electronic structure of USb$_2$. To consider the magnetic ordering in the strong spin-orbit coupling $5f$-electron system, (j, m$_j$) basis set was used without any orbital symmetries. Note that the AFM breaks not only the combination of time-reversal and discrete lattice translation, but also the rotational symmetry. However, because there is strong spin-orbit coupling that is included in the DMFT calculation, the easy axis is fixed and the computation needs to be done in that reference system. Furthermore, it is the rotational symmetry of coupled spin and orbital degrees of freedom, which is broken in the AFM phase.

The crystal field effect was not taken into account for simplicity. The crystal field splitting is ~ 10 meV, which is quite smaller than the spin-orbit splitting (~ 1 eV) and the exchange splitting (~ 100 meV). Hence, the crystal field effect does not change the physics in this study.

We find that the electronic structure calculated in DMFT is qualitatively consistent with STM data. At T = 100 K, two broad peaks at ~ - 60 meV and ~ + 70 meV are well captured in DMFT. The peak at ~ - 60 meV has, primarily, (5/2, ±5/2) and (5/2, ±3/2) orbital characters, while the peak at ~ + 70 meV has (5/2, ±1/2) orbital characters. Upon cooling, the sharp Kondo resonance peak develops near the Fermi level and has (5/2, ±1/2) orbital characters. On the other hands, the other orbitals, namely (5/2, ±5/2) and (5/2, ±3/2), give the magnetic moment. Orbital differentiation is clearly shown in the DMFT calculations: (5/2, ±1/2) orbitals show Kondo physics and (5/2, ±5/2) and (5/2, ±3/2) orbitals show magnetism governed by RKKY physics. The calculated DMFT electronic structure is shown in figure S4.

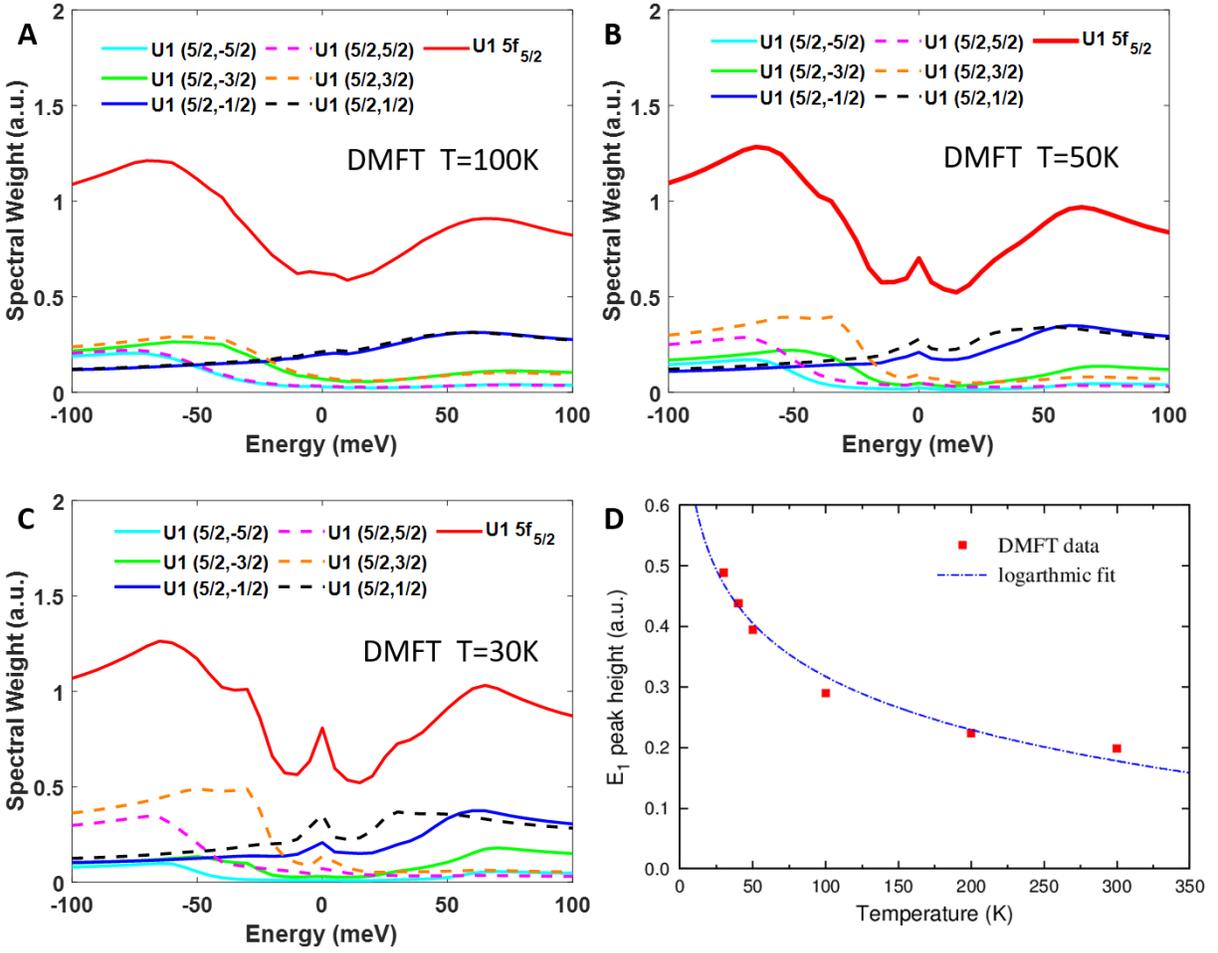

**Fig. S4. DMFT:** (A,B,C) DFT-DMFT calculated electronic structures (red line) and their deconvoluted orbital components underneath at T=100K, 50K & 30K, respectively. (D) Temperature dependence E1 peak height from DMFT. The line is a logarithmic fit to the data.

## S5. *dI/dV* spectra fitting to Fano-lineshape

Figure S5 demonstrates the fitting process for the *dI/dV* spectra of figures 2 and 3 of the main text. The data at low temperatures (T < 45 K) are fitted to the sum of three Fano lineshapes representing the three resonances at $E_{1,2}$ and $E_0$ with a 2$^{nd}$ order polynomial background (see Fig.S5a). The Fano lineshape is described as

$$\frac{dI}{dV} \sim A \frac{(\frac{V-E}{\Gamma} + q)^2}{1 + (\frac{V-E}{\Gamma})^2}$$

Where E characterizes the resonance energy, Γ the resonance width (HWHM) and  *q* the tip-sample coupling, also known as the asymmetry parameter of the Fano lineshape. A is related to the amplitude of the resonance. All the fits are presented in Fig.2a of the main text in excellent agreement.

For the high temperature data (T>45K), only the data inside the AFM gap where fitted to a single Fano lineshape and a 2$^{nd}$ order polynomial background (see Fig.S5b).

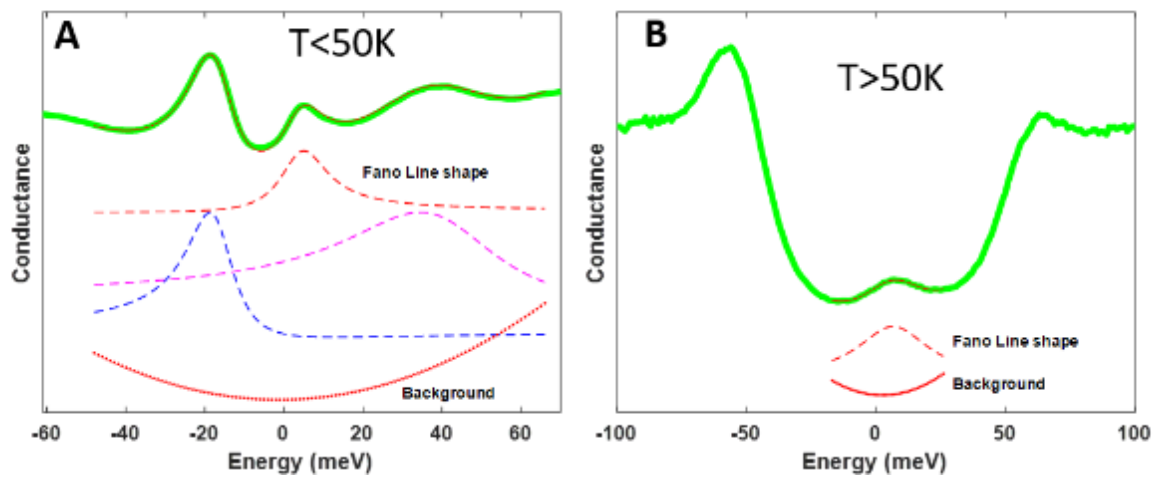

**Fig. S5. dI/dV Fano fit: (A)** (T<50K) The data are fit to 3 Fano line shapes (red dashed line), one for each peak, and a 2$^{nd}$ order polynomial background (red dotted line). **(B)** (T>50K) The data are fit to 1 Fano line shape (red dashed line) and a 2$^{nd}$ order polynomial background (red dotted line). The red solid line represents the overall fit to the data (green solid line).

The height of the peaks in figure 2b,c are not be extracted directly from the Fano equation. The height of the resonant peak depends on both the amplitude of the Fano line and the parameter q through the relation $A(1 + q^2)$. This expression though propagates the error of $a$ and $q$ creating a large uncertainty in the overall height on the peak since both parameters depend on each other. Instead, the heights presented in fig.2b, c are directly extracted from the maxima of the peaks relative to the lowest point in every dataset.

# S6. Spatial modulation of the Kondo hole

The same fitting procedure demonstrated in S5 has been used to extract the spatial effect of the Th atoms on the Kondo lattice described in Fig.3 of the main text. Here we also show the parameters for the $E_2$ resonance.

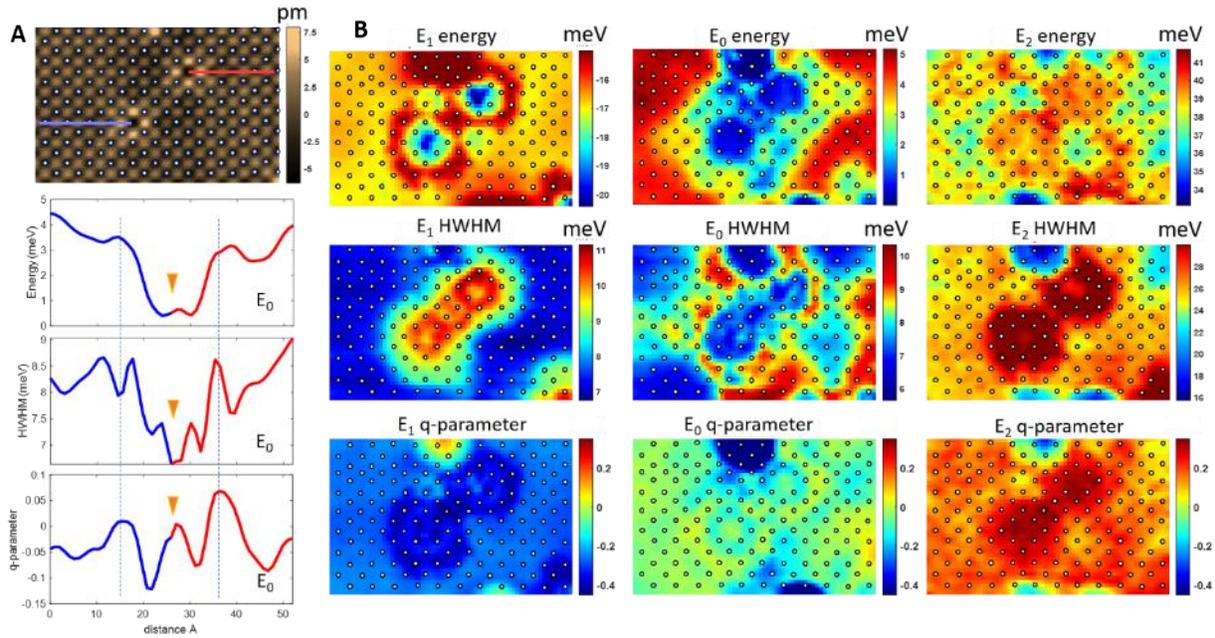

**Fig. S6. Spatial modulation of Kondo hole:** Spatial effect of the Th-dopants on the Kondo lattice. (A) topograph. (B) Energy, HWHM and q-parameter of $E_1$, $E_0$ and $E_2$ for the same map as in figure 3 of the main text. Note that the q-maps have the same color scale throughout the three maps.

The above maps clearly show spatial oscillatory behavior of the energy, width, and asymmetry of the resonances. For the $E_0$ resonance, we further show linecuts across the two Th-dopants (blue and red lines in the topograph). Linecuts along the opposite or vertical directions is obscured by interference between the two Th-dopants and a dimer defect on top of the map, respectively. All parameters show the oscillatory behavior. The spatial variation of the HWHM further shows two periodicities, a small wavelength corresponding to the atomic lattice along the ($\pi/a$, $\pi/a$) direction and a longer wavelength (which acts as an envelope) that matches with the oscillations in the Energy and q-parameter (shown with vertical dashed lines). The wavelength of these oscillations corresponds to about 12.5 Å and relates to a wave vector of q ~ 0.47 along the ($\pi/a$, $\pi/a$) direction. Note that the oscillation wavelength in this case is extracted from a single dopant without enough statistical average to make a reliable statement.

The maps of the tip-sample coupling parameter q for $E_1$, $E_0$ and $E_2$ are plotted with the same color-scale. It becomes clear that the values of q for the 3 energies are different. Furthermore, their oscillatory behavior also varies as shown by the linecuts across the Th-dopant in figure S7. This suggests that the STM tip is coupled to different orbitals when probing these resonances supporting the concept of orbital selectivity.

The oscillatory behavior of the hybridization strength has been predicted theoretically around the Kondo hole (ref 37 & 38 of main text). Experimentally it is realized through the oscillation of the peak width and asymmetry parameter q. Fig S8 shows the hybridization of the f and spd bands and the two band gaps that open, the direct bandgap $\Delta_d$ and the indirect bandgap $\Delta_i$. Each flat band manifests as a peak in the DOS. Depending on the magnitude if $\Delta_d$, and the small dispersion of the f-band, the peaks can be pushed away from each other resulting in the dual peaks seen in fig S8 or overlap giving a single peak as happens in our data for $E_0$. In the vicinity of the Kondo hole, the magnitude of the hybridization, and therefore the bandgap, oscillates manifesting itself in the oscillatory behavior of the peak width. Similarly, the asymmetry parameter q, which is defined as the tip's coupling ratio to f-electrons (Kondo resonance) and conduction electrons, is expected to oscillate as the resonance amplitude oscillates in real space.
An oscillatory pattern for the energy around the Kondo hole has not been explicitly predicted but is nonetheless important to report. Two reasons can be behind the oscillatory shift in energy seen in Fig.S1D. The first is the local tiny change in the overall valence of the U-atom (or the missing U atom at the Th substitution cites) that is known to shift the energy of the Kondo resonance (ref. 6). The other could be a result of the spatial oscillatory behavior of the c and f electron local DOS.

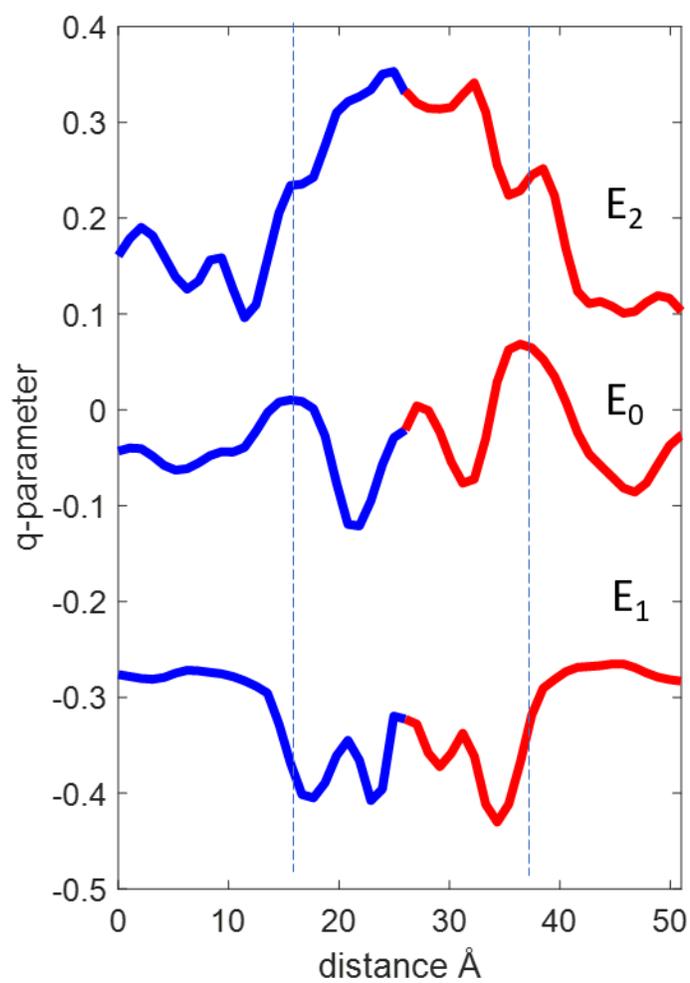

**Fig. S7. Linecuts of q-maps:** Linecuts across the Th-dopant of $E_1$, $E_0$ and $E_2$ from the q-parameter maps. The tip-sample coupling parameter exhibits different oscillatory behavior across the three energies.

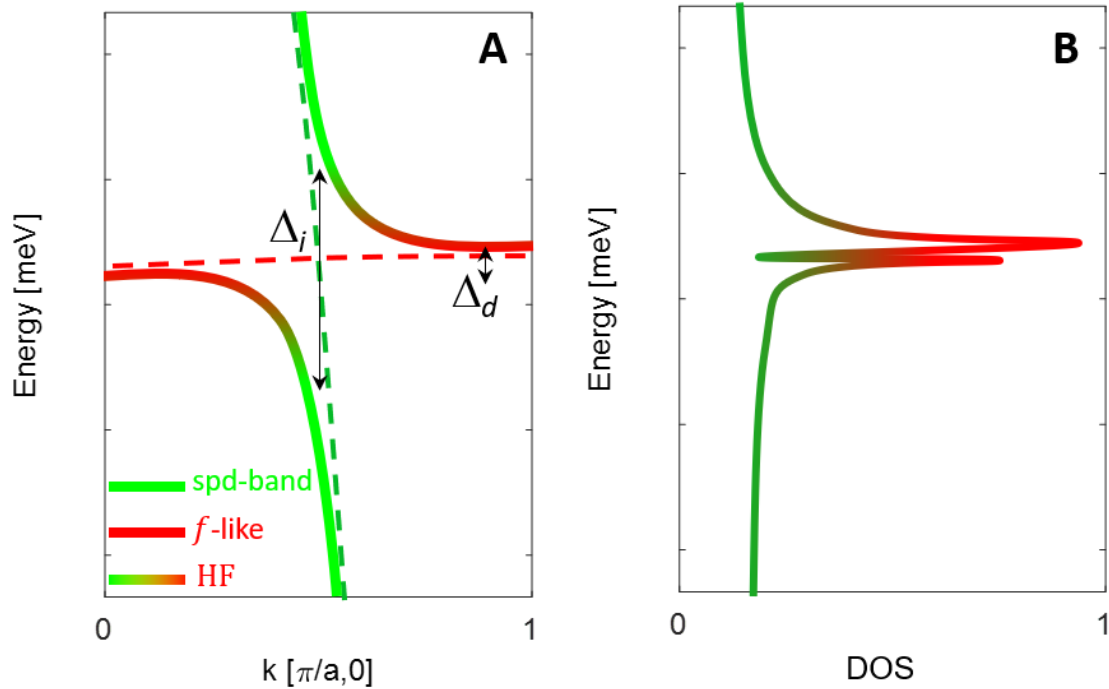

**Fig. S8. Hybridized bands and DOS: (A)** Schematic of the heavy fermion band structure with the direct and indirect gaps. The dashed lines correspond to the unhybridized spd (green) and f (red) bands, whereas the solid lines correspond tot eh hybridized bands. **(B)** Schematic tunneling density of states showing enhancement at energies corresponding to the flat bands.

We further used a different method to study the spatial behavior of the different parameter of the Kondo-hole. Using the conductance map of figure 4 of the main text, we fit the spectra at every pixel (256 x 256 pixels) to the Fano lineshape model discussed in S5. The large size of the map would not allow to distinguish the local modulation around each Th atom in real space, but the effect should manifest itself in the Fourier Transform of the Fano parameter maps. Figure S9 shows the FT of the extracted energy, HWHM and q-parameter for the $E_0$ peak. We observe a weak signal in all three FTs. A weak feature is observed near q ~ 0.5 along the ($\pi/a, \pi/a$) direction. While one would expect a much stronger signal in FT due to the contribution of all Th-dopants, the signal is considerably weak due to the fact that Th dopants from layers below (substituting in the 2$^{nd}$ and 3$^{rd}$ U-layers) also contribute to the surface conductance. Since the effect of these defects (2$^{nd}$ and 3$^{rd}$ layer U substitution) will have a different phase shift with respect to those right below the surface (1$^{st}$ layer U substitution) (see Fig. S1(D)), the different signals interfere with and obscure the overall signal.

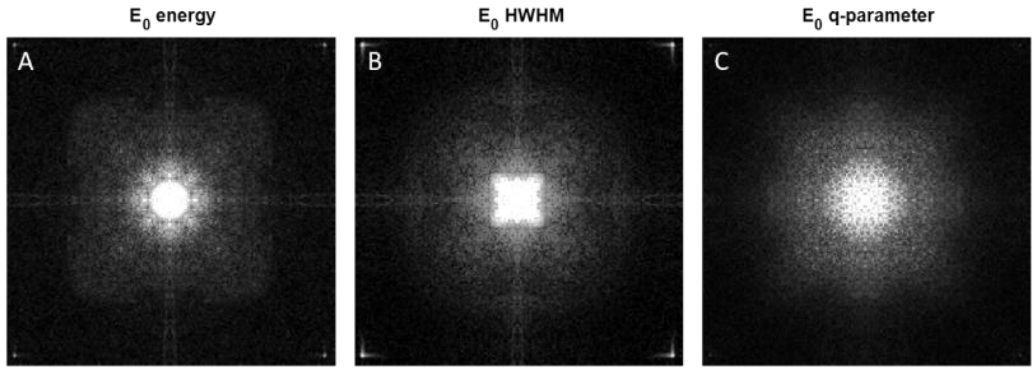

**Fig. S9. FT of fit parameters: (A)-(C)** FT of the energy, HWHM and q-parameter from the Fano fit, described in section 3, to the conductance map of figure 4 of the main text. The figures have been 2-fold symmetrised to enhance the signal.

## S7. Energy-Momentum structure from quasiparticle interference

Figure S10 shows the discreet Fourier transform of the conductance maps on a different sample with x=0.5% Th doping. The data have been 4-fold symmetrised to enhance the signal. We observe the same quasiparticle interference (QPI) features and their energy evolution, consistent with that observed for x=0.3% in fig. 4 of the main text.

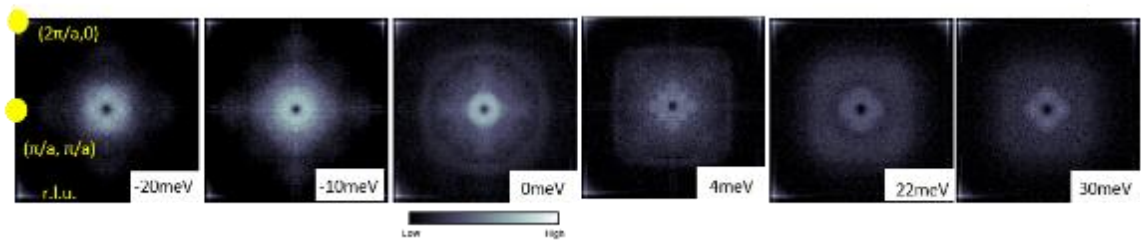

**Fig. S10. FT of conductance map for x=0.5%:** Discreet Fourier transform of the conductance map for x=0.5%. The data have been 4-fold symmetrised.

The symmetrisation process serves only to enhance the strength of the QPI features. Figure S11 shows selected pairs of the raw FT data and their symmetrized counterparts for selected energies. The same features are visible in each pair. The large-scale inhomogeneity (long wavelength) that results in the intense peak at the center (q ~ 0) of the FTs are subtracted by a Gaussian-filter. This results in the dark spot seen in the centre of the Fourier transforms and has no effect on the QPI signal.

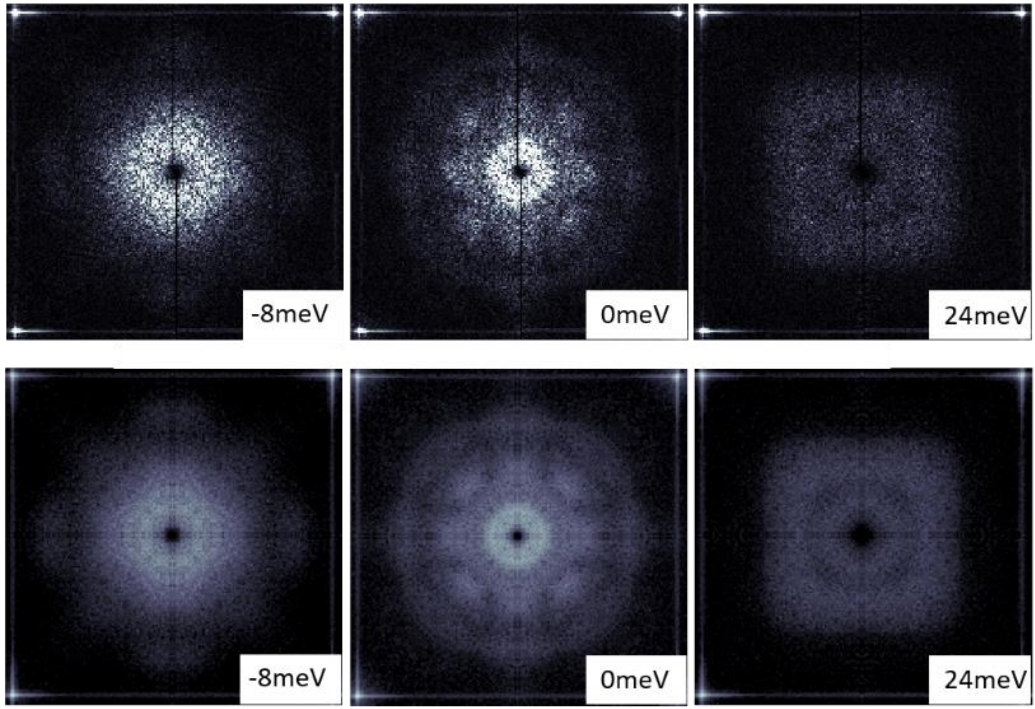

**Fig. S11. Raw-symmetrised data comparison:** Pairs of raw (top) and symmetrised (bottom) data for a few selected energies.

## S8. Extraction of the QPI features

It is important to accurately extract the momenta of the QPI features from the FTs of the conductance maps. Figure S12(A), (B) show line cuts of the FTs taken along the ($\pi/a$, $\pi/a$) and ($2\pi/a$, 0), respectively, at different energies (blue dotted lines). The data are fit to two Gaussians with a 2$^{nd}$ order polynomial background (red lines). The extracted features are plotted in Fig.5 of the main text as datapoints on top of the color-scale FT linecuts. For the data in Fig. S12(A) one can clearly observe two dispersive bands denoted by the orange and black triangles. The orange band disperses rapidly (broadens) near -18 meV signaling its hybridization with the $E_1$ band. The black band disperses rapidly around 4meV due to its hybridization with the $E_0$ band. On the other hand, in Fig. S12(B) we observe a single band that becomes very broad near 6meV due to its hybridization with the $E_0$ band, as observed in fig. 5 of the main text. For this particular direction and at those energies, a Gaussian fit cannot mathematically represent the data and for this reason we do not try to fit those linecuts.

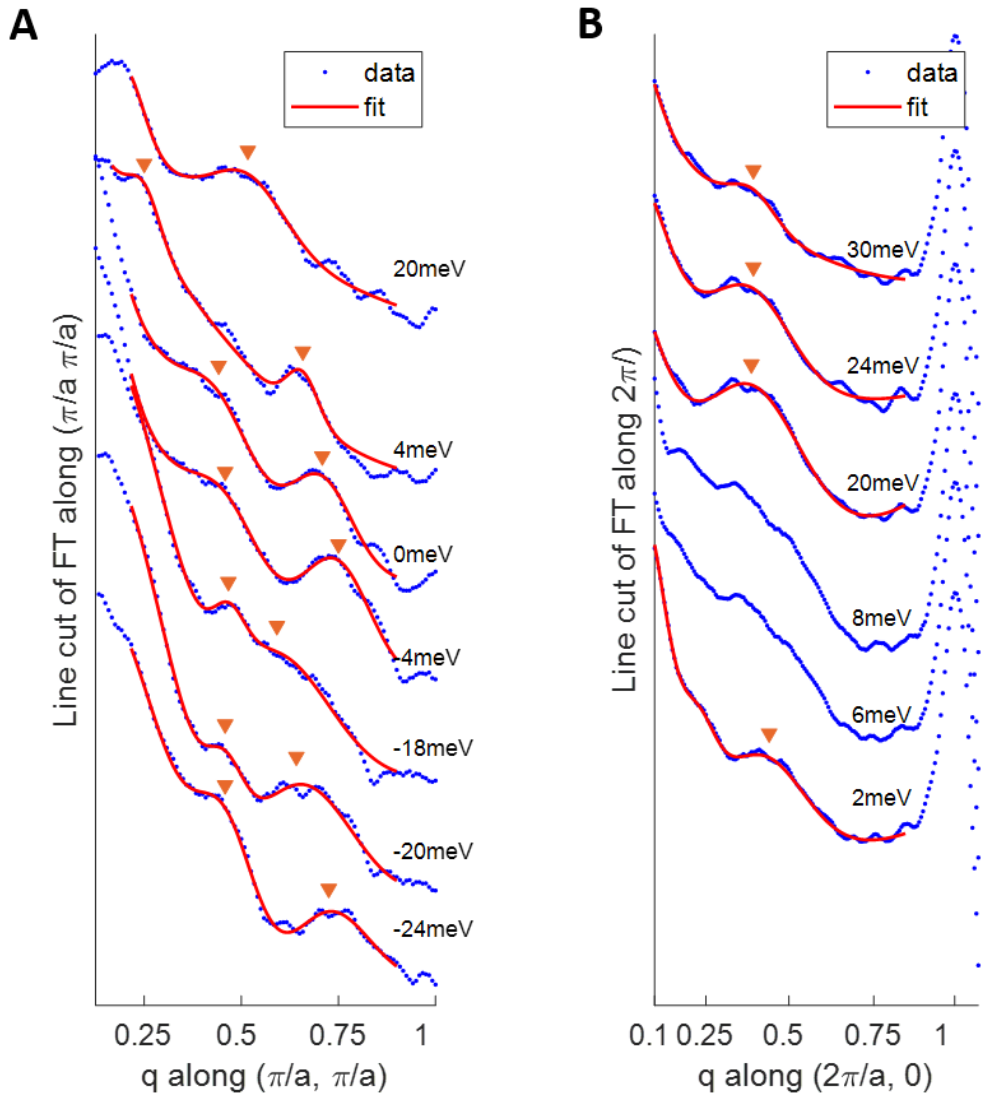

**Fig. S12. Linecuts across QPI features:** Linecuts and their fits for different energies along the **(A)** (π/a, π/a) and **(B)** (2π/a, 0) direction. The lines are shifted upwards for clarity. The arrows indicate the QPI features as they appear in the FT linecuts.